\documentclass[final]{svjour3}
\usepackage{graphicx}
\usepackage{rotating}
\usepackage{amssymb}
\usepackage{mathptmx}
\usepackage[sort&compress,numbers]{natbib}
\usepackage{authblk}
\usepackage[utf8]{inputenc}
\usepackage{textcomp}
\usepackage{upgreek}
\makeatletter
\journalname{Journal of Low Temperature Physics}

\date{November 4, 2017}

\author{B.A. Steinbach \and J.J. Bock \and H.T. Nguyen \and R.C. O'Brient \and A.D. Turner}

\institute{B.A. Steinbach \and J.J. Bock \at Department of Physics, California Institute of Technology, Pasadena, California 91125, USA
\and
H.T. Nguyen \and R.C. O'Brient \and A.D. Turner \at Jet Propulsion Laboratory, Pasadena, California 91109, USA
}

\begin{document}

\newcommand{\hdblarrow}{H\makebox[0.9ex][l]{$\downdownarrows$}-}
\title{Thermal kinetic inductance detectors for ground-based millimeter-wave cosmology}

\maketitle

\begin{abstract}

We show measurements of thermal kinetic inductance detectors (TKID) intended for millimeter wave cosmology in the 200-300 GHz atmospheric window.
The TKID is a type of bolometer which uses the kinetic inductance of a superconducting resonator to measure the temperature of the thermally isolated bolometer island.
We measure bolometer thermal conductance, time constant and noise equivalent power.
We also measure the quality factor of our resonators as the bath temperature varies to show they are limited by effects consistent with coupling to two level systems.


\end{abstract}

\section{Introduction}

Measurements of the cosmic microwave background (CMB) polarization in search of a gravitational wave signal from the epoch of inflation are limited by a foreground of galactic dust \citep{keck_array_and_bicep2_collaborations_improved_2016}.
Improving constraints on r, the tensor-to-scalar ratio, hinges on imaging galactic dust in the 200-300 GHz atmospheric window, where the brightness of dust relative to the CMB is enhanced relative to 95 GHz or 150 GHz \citep{kamionkowski_quest_2016}.
In this paper we explore the thermal kinetic inductance detector (TKID) as a
path to fill a 200-300 GHz focal plane, inspired by detector developments for
x-ray spectroscopy \citep{ulbricht_highly_2015,quaranta_mitigation_2014,miceli_towards_2014,cecil_optimization_2015,lindeman_resonator-bolometer_2014}.
TKIDs are bolometers whose thermometer exploits the temperature dependence of the kinetic inductance effect.
As in a direct absorber kinetic inductance detector (KID) \citep{day_broadband_2003,mccarrick_horn-coupled_2014,dober_optical_2015}, the resonant frequency of an LC resonator shifts in response to the quasiparticle density in a superconducting inductor.
However, in a TKID, rather than directly breaking pairs, photons are absorbed on a suspended island shared by the inductor and quasiparticles are produced thermally.
Like KIDs, TKIDs can be frequency multiplexed by assigning each detector a different resonant frequency, and weakly coupling the resonators to a shared readout transmission line.
\\\indent The potential advantage of TKIDs is engineering freedom.
In a KID, the function of electromagnetic absorption, conduction of optical power out of the detector and to the bath, and low frequency readout are performed by the kinetic inductor which must be simultaneously optimized for all three functions.
In a TKID, these functions can be separated into a load resistor, a silicon nitride membrane, and superconducting inductor, which can be independently optimized, at the cost of a many layer fabrication process.
Fortunately, we can leverage the many similarities this process has to existing transition edge sensor (TES) bolometer fabrication \citep{ade_antenna-coupled_2015}.

\section{Device design}

\begin{figure}[!tbp]
\centering
\begin{tabular}{ccc}
	\includegraphics[width=0.3 \textwidth]{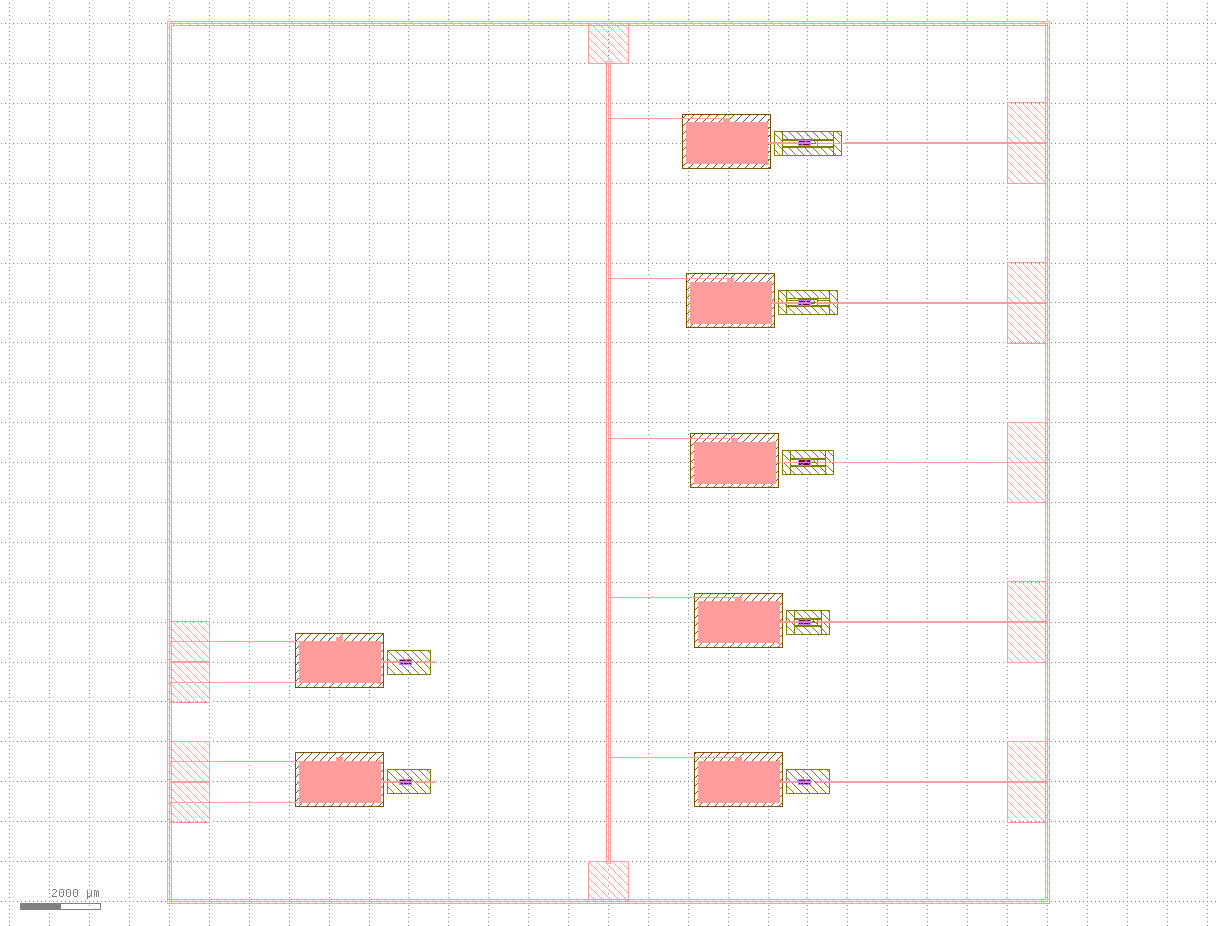}
	&
	\includegraphics[width=0.3 \textwidth]{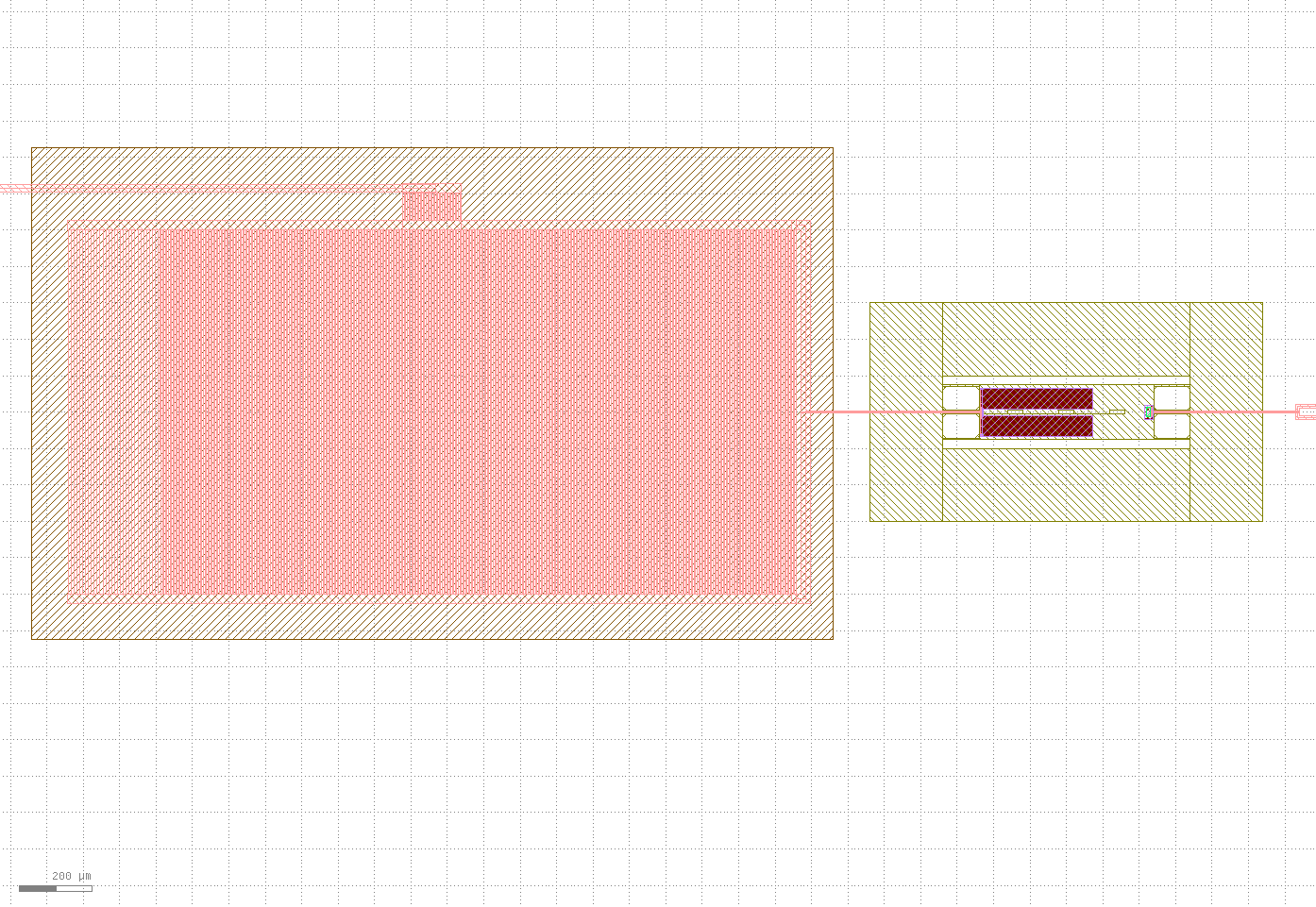}
	&
	\includegraphics[width=0.3 \textwidth]{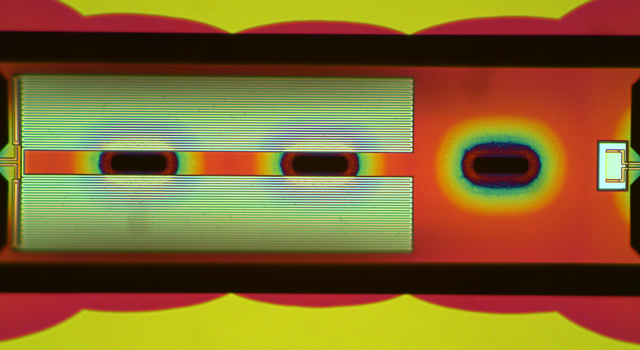}
\end{tabular}
\caption{Left, layout of TKID chip mask.
The readout line runs vertically down the center of the chip.
Five active resonators on the right side of the chip have heater resistors that run to wirebond pads on the right edge of the chip.
The two devices in the lower left of the chip are resistance and $T_c$ test structures for the aluminum inductor layer.
Middle, zoom in on one resonator of TKID chip mask, showing the resonator capacitor on the left and island on the right.
Right, photograph of one bolometer island, showing the meandered aluminum inductor on the left and the gold heater resistor on a niobium washer on the right.
Color banding around the bolometer release holes suggests the oxide etch stop has been consumed and up to $\approx$100 nm of the silicon nitride etched by the xenon difluoride.}
\label{fig:mask_all}
\end{figure}

We fabricated a test chip to study the suitability of TKIDs for ground based observations.
Images of the mask and fabricated device are shown in Fig.~\ref{fig:mask_all}.
The test chip contains 5 TKIDs, one with an unreleased bolometer, and four with bolometer leg lengths 100 µm, 200 µm, 300 µm 400 µm. 
The bolometer geometry is based on the design used for the Keck/BICEP program, which uses six parallel legs to mechanically suspend and thermally isolate a silicon nitride island \citep{kuo_antenna-coupled_2008}.
To facilitate step coverage, the mechanical substrate of the bolometer island, low stress silicon nitride, is 1/3 the thickness of that used for the Keck/BICEP TES bolometers, so we fabricated multiple leg lengths to measure the thermal conductivity (G) as a function of leg length for the thinner silicon nitride.
The temperature sensing inductor on the island is a 16 mm long, 1µm wide meandering trace of 50nm thick aluminum.
On the island, niobium traces contact the inductor leads and run off along the island legs to an interdigitated resonator capacitor.
The silicon nitride layer under and around the capacitor is removed so that the resonator capacitor is deposited directly on the silicon wafer.
To vary the resonant frequencies from 260 to 315 MHz, the number of fingers in the resonator capacitors is varied.
\\
\indent The resonators are weakly coupled to a readout transmission line through a second small interdigitated capacitor on one side of the resonator, and through the parallel plate capacitance of the resonator to the ground of the chip holder to complete the circuit.
The parallel plate capacitance to ground acts on both sides of the resonator, and on the coupling capacitor side acts to weaken the coupling.
This was not accounted for in the design, so the devices are more weakly coupled than intended with coupling $Q_c$=80,000-140,000 rather than 20,000.
\\\indent
Rather than integrate an antenna directly into this chip, power is deposited onto the bolometer island via a resistive heater and DC current source.
This calibrates the leg thermal conductivity and detector noise directly in units of watts.
On the tested chip, two released bolometers (100 µm and 400 µm legs) and the unreleased bolometer were fully functional.  Of the remaining two bolometers, one resonator was broken and one heater was left disconnected from the cryostat wiring.
\\\indent
The devices were fabricated on a high resisivity silicon substrate $\approx$500 µm thick.
A 40 nm silicon dioxide and 300 nm low stress silicon nitride film are deposited on top of the wafer, which form the membrane layer.
E-beam evaporated lift-off aluminum 50 nm for the inductor is deposited on top of this.
The gold resistor layer is 200 nm thick, e-beam evaporated, and patterned with lift-off.
Next, the silicon nitride and silicon dioxide are etched to expose silicon and define the island and the region where the capacitor will be deposited.
The niobium layer for the capacitor and wiring is 400 nm thick and patterned with lift-off.
Finally, holes in the island are drilled by deep reactive ion etching and the island released with xenon difluoride.

\section{Device Characterization}
\begin{figure}
\centering
\begin{tabular}{cc}
\includegraphics[width=2.0in]{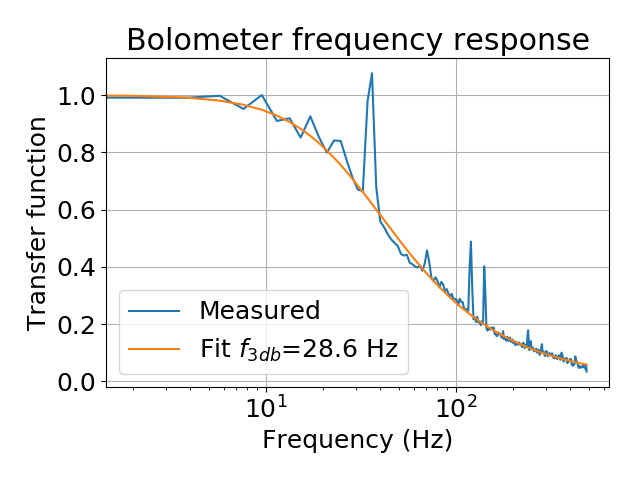}
\end{tabular}
\caption{Measured frequency response of the 400 µm leg bolometer.  Blue shows the measured resonator frequency power spectrum response to a log-swept chirp heater excitation, normalized to the power spectrum of the log sweep.  Orange shows the one pole model fit.}
\label{fig:timeconstant}
\end{figure}

The film properties of the aluminum are measured via a test structure on one side of the chip consisting of niobium leads running to an aluminum inductor identical to the inductor on the bolometer islands.  
The transition temperature of the aluminum film is 1.32 K and the sheet resistance is 0.23 $\Omega/\Box$, which implies a thin film Mattis-Bardeen kinetic inductance of $0.24 pH/\Box$ \citep{zmuidzinas_superconducting_2012}.
The niobium to aluminum contact is zero resistance up to the critical current of $\approx$100 µA.
\\\indent
To measure the bolometer thermal conductance to the bath, the resonator is first calibrated by sweeping the bath temperature while measuring the resonant frequency with a small readout excitation of -110 dBm (0.01 pW).
Then the bath is returned to base temperature and the island heater power is swept while measuring the resonant frequency to measure the island temperature.
The measured conductances are 26 pW/K and 64 pW/K for the 400 µm and 100 µm detectors respectively, at an island temperature of 0.38 K.
0.38 K operating temperature compromises between quality factor ($Q_{i,MB} \approx 10^4$) and noise equivalent power (NEP) in an ideal aluminum TKID.
\\\indent
To measure the responsivity as a function of frequency of the 400 µm leg bolometer, a function generator supplied a DC offset and a sine wave chirp with a logarithmic sweep in frequency to the island heater.
The frequency of the resonator was monitored using the technique described in Section \ref{sec:noise}.
The power spectrum of the response is shown in Fig.~\ref{fig:timeconstant}.
The response is a good fit to a one pole model with 3 dB frequency 28.6 Hz or a heat capacity of 0.14 pJ/K.

\section{TLS}

The power and temperature dependence of the resonator quality factors show substantial loss consistent with the the behavior of two level systems (TLS).
We follow standard methods to model the resonator and the temperature and power dependence of TLS\citep{mccarrick_horn-coupled_2014,gao_experimental_2008,von_schickfus_saturation_1977}.
A model for $S_{21}$ of a resonator is fit to extract the resonator quality factor $Q_r$ and complex coupling quality factor $Q_e$.
The internal resonator quality factor is estimated from the difference of $Q_r$ and the real part of $Q_e$.
The internal quality factors are then fit to a sum of contributions from TLS (Eq. \ref{eq:tls}), Mattis-Bardeen quasiparticles from the measured $T_c$, and a constant $Q_i$ for loss due to other sources.
Power dependence is captured by the terms $P_g$, the generator power, and $P_c$, the generator power above which TLS is saturated.

\begin{equation} Q_{TLS} = Q_{TLS_0} \frac{\sqrt{ 1 + \frac{P_g}{P_c} }}{\tanh \left( \frac{ h f}{2 k T} \right) } \label{eq:tls} \end{equation}

\begin{figure}
	\centering
	\begin{tabular}{cc}
	\includegraphics[width=2.0in]{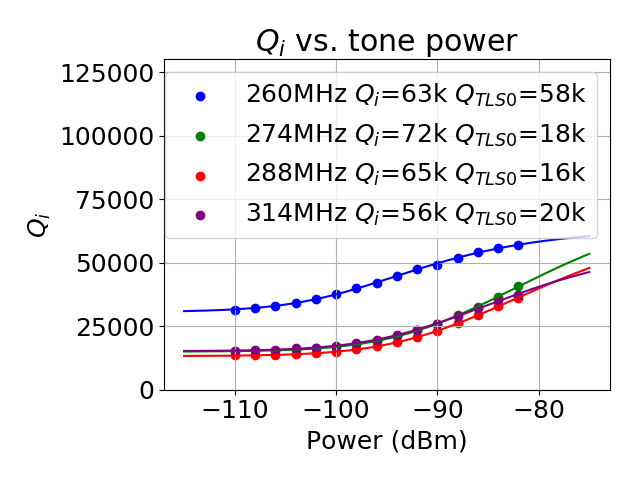}
	\includegraphics[width=2.0in]{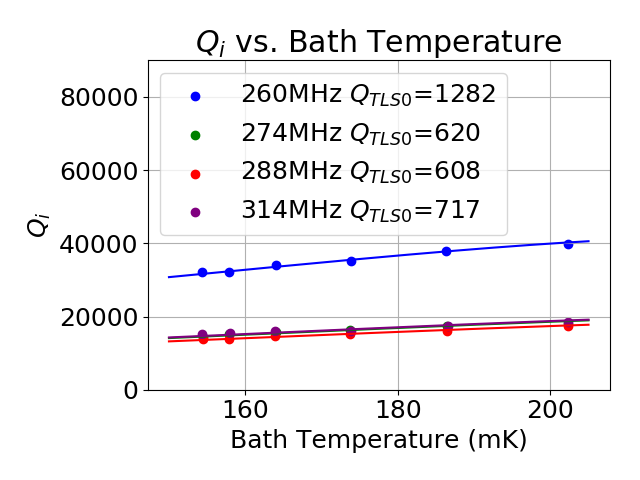}
	\end{tabular}
	\caption{Power and temperature dependence of $Q_i$.  Power dependence is measured at 0.15 K while temperature dependence is measured at -110 dBm.  Points are data and lines are fits.  Legend shows fitted parameters.  The blue curve with twice the $Q_i$ of the other devices is the unreleased device.}
	\label{fig:ptvq}
\end{figure}

As shown in Fig.~\ref{fig:ptvq}, $Q_i$ is strongly limited by a TLS consistent loss where $Q_i$ increases with power and temperature.
The limiting performance in the unreleased devices may be due to side wall coating during the niobium lift-off step, visible as flags of metal or resist along the lines of the capacitor.
Subsequent unreleased devices fabricated using etch-back to pattern the niobium capacitor show low temperature and power $Q_i > 60,000$.
\\\indent
The released devices show much lower $Q_i$.
We are currently considering two possibilities.
Chemical damage is one possibility, as the long distance ($\approx$100 µm) between release hole and edge of the island necessitated a long exposure to xenon difluoride (15 pulses, 45 seconds per pulse, 3 torr).
The island can be re-engineered to reduce this distance to 15 µm.
A second possibility is mechanical damage due to stress on the island during and after the release process.

\section{Noise}
\label{sec:noise}

\begin{figure}[!tbp]
\centering
\begin{tabular}{cc}
	\includegraphics[width=0.5 \textwidth]{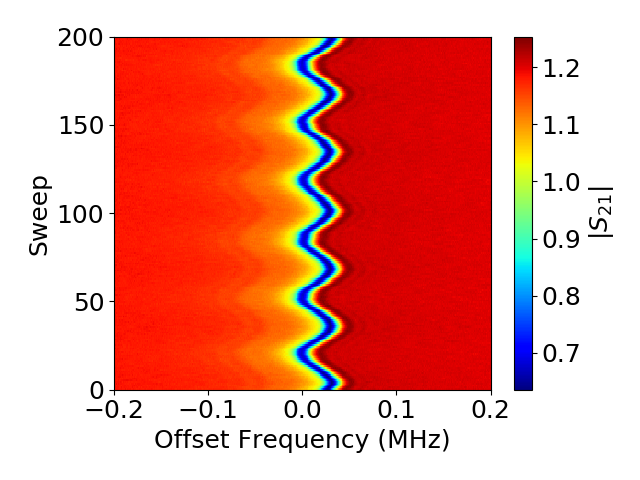}
	&
	\includegraphics[width=0.5 \textwidth]{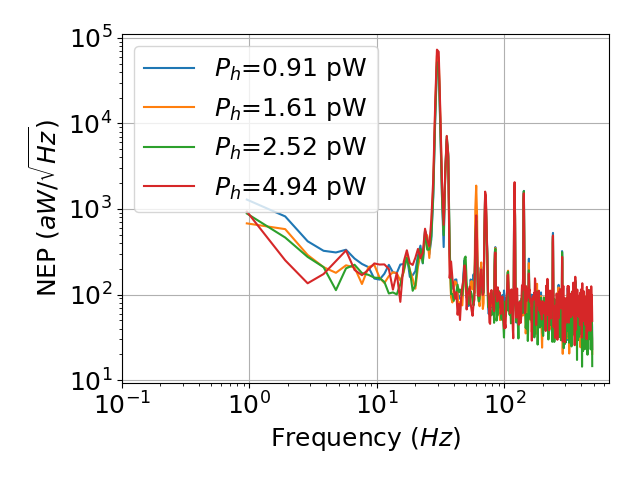}
\end{tabular}
\caption{Left, Waterfall plot of network analyses from chirp readout.
Each horizontal slice is a single network analysis, separated by the one below it by one millisecond.
The TKID's heater is excited by a 30Hz sine wave with a DC offset of 1.6pW and an of RMS 0.3pW, sufficient to move the heater several resonance widths.  This is an exaggeration of the motion used to perform NEP calibration and demonstrate the capability to track the resonator across large changes in power.
Right, NEP of 400 µm leg TKID measured at several heater powers.
Amplitude spectral density is not corrected for thermal time constant.
The large spike at 30 Hz is the NEP calibration tone.
The neighboring spike at 35 Hz is due to electrical pick up from the pulse tube valve stepper motor.
	}
\label{fig:chirps}
\end{figure}

To characterize NEP, the resonator frequency is tracked by continuous network analysis.
Once per millisecond, a chirp generated by a software radio sweeps in frequency across a 1 MHz bandwidth centered on the resonant frequency of one detector, from high to low frequency.
A 1 ms chirp is slow compared to the time constant of the resonator ($\approx$10 µs) but fast compared to the time constant of the island ($\approx$5 ms).
The ratio of the fourier transforms of the received signal and input chirp is proportional to the transfer function $S_{21}$.
The position of the resonance in each transfer function sweep is fit to produce a resonator frequency timestream sampled at 1 kHz.
Fluctuations in gain and phase of the readout are removed by fitting a baseline of the transfer function away from the resonance.
The wide bandwidth of the sweep compared to a single tone readout allows the resonator to be tracked across large temperature changes corresponding to frequency shifts of many resonance widths.
We find this a practical technique for reading out and characterizing single resonators, but for chirp rates slow compared to the resonator ring down time, a time multiplexing penalty makes it impractical for large numbers of resonators.
An example of the continuous network analysis from the chirp readout scheme is shown in the left panel of Fig.~\ref{fig:chirps}.
To measure the NEP of the detector, the resonator frequency is monitored while calibrating resonator frequency into power by applying a sine wave of known frequency and power to the island heater.
The resulting noise power spectrum is shown in the right panel of Fig.~\ref{fig:chirps}, and reaches $\approx$200 aW/$\sqrt{Hz}$.
This is a few times the single mode photon noise under the south pole atmosphere, and an order of magnitude larger than the phonon noise expected for the 0.38 K island temperature.
Despite the large contribution to detector $Q_i$ from TLS loss, the excess does not appear to originate from TLS, as it is highly correlated between different resonators.
We suspect environmental effects such as pulse tube vibrations, the magnetic environment \citep{flanigan_magnetic_2016} and the infrared environment \citep{barends_minimizing_2011} are responsible, and a study of those effects is on going.

\begin{acknowledgements}
This work was supported by JPL's Research and Technology Development Fund for projects R.17.223.057 and R.17.223.058.
\end{acknowledgements}

\pagebreak

\bibliographystyle{unsrtnat}
\bibliography{resobololtd2017}

\end{document}